\documentstyle[12pt]{article}
\begin{document}
\title{Fluctuations in the Quantum Vacuum}
\author{B.G.Sidharth \\
B.M.Birla Science Centre,Hyderabad, 500463,India}
\date{}
\maketitle
\begin{abstract}
We argue that it is a fluctuational underpinning of the Quantum vacuum which
on the one hand gives a stochastic character to the conservation laws, and
on the other is required for explaining the recently observed acceleration
of the universe. This also provides us with the arrow of time, in a
consistent cosmology.
\end{abstract}
\section{Introduction}
Classical Physics is based on a deterministic spacetime. It is also true that
the equations of Classical Physics are time reversible. The equations of
Quantum Mechanics too are time reversible, though here it is the probabilities
which are determinable. Irreversible processes are phenomenological, according
to this thinking. That is irreversibility arises due to the role of the
observer.\\
One important difference which is brought about by Quantum Theory is that the
smaller the spacetime intervals we consider, the greater the Uncertainity
in the energies and momenta wherein the Uncertainity in energy results in the
creation and destruction of particles. As Feynman put it\cite{r1}, "created
and annihilated, created and annihilated - what a waste of time". The Quatum
vacuum is thus an arena of frenzied activity at the micro scale. Nevertheless,
conservation of energy is respected. Particles may appear to be created, but
this energy is returned to the Quantum vacuum within the Uncertainity time
interval. There are wild incessant fluctuations in energy, but these miraculously
cancel out on the average. Energy is conserved on the average.\\
However all of this still respects a smooth spacetime manifold, what Witten
has termed "Bosonic spacetime"\cite{r2}.\\
\section{Non Commutative Spacetime}
Latest studies for example Quantum SuperString Theory or the author's stochastic non commutative
spacetime, take us beyond the above approximate description to what Witten
calls Fermionic spacetime\cite{r2,r3,r4,r5}. In this case two spacetime
coordinates like $x$ and $y$ do not commute but rather we have the equation
\begin{equation}
[x,y] \approx 0(l^2)\label{e1}
\end{equation}
and similar equations, where $l$ defines the Compton scale.\\
Another way of looking at (\ref{e1}) is that the usual Uncertainity Principle is
modified:
\begin{equation}
\Delta x \cdot \Delta p \sim \hbar + h',\label{e2}
\end{equation}
$$h' \sim \frac{l^2 \cdot (\Delta p)^2}{h}$$
This is an expression of the well known duality\cite{r2,r6}, which provides an
interface between the micro world and the macro universe. Infact $h'$ defines
this interface. We can now show that $h'$ results in a non conservation of
energy. Infact from (\ref{e2}) we have
\begin{equation}
\Delta p \sim \frac{h R}{l^2}\label{e3}
\end{equation}
where $R =\Delta x$ is the radius of the universe.\\
We have an equation for energy similar to (\ref{e3}) also. Using the well
known Eddington formula, $R \sim \sqrt{N} l$, where $N \sim 10^{80}$ is the
number of particles in the universe, (\ref{e3}) becomes
$$\Delta p \sim \sqrt{N} mc$$
and the similar equation for energy is
\begin{equation}
\Delta E \sim \sqrt{N} mc^2\label{e4}
\end{equation}
An interpretation of (\ref{e4}) follows naturally from the model of fluctuational
cosmology\cite{r7,r8}. Here the $\sqrt{N}$ particles are created fluctuationally
out of the Quantum vacuum, and equation (\ref{e4}) gives the energy of these
particles. It must be emphasized that the above cosmological model provides
an explanation for the otherwise miraculous large number coincidences, apart
from predicting an ever expanding, accelerating universe, as indeed the
latest observations confirm.\\
It must also be emphasized that in the above
model the particles are created unidirectionally, that is there is no
destruction of particles. It is in this sense that the above model goes
beyond the conventional conservation laws, as symbolised by the extra
Uncertainity term $h'$ in equation (\ref{e2}). Infact as argued elsewhere\cite{r9},
the conservation laws themselves are not iron clad, but are stochastic in
nature. There is a certain resemblance of this model to Dirac's large number
cosmology\cite{r10}, wherein also particles are created. However this latter
model has some inconsistencies. What is interesting is that Dirac, on the
one hand brilliantly hypothesized that the large number relations like the
Eddington formula referred to earlier were not mere accidents, but rather
were symptomatic of a profound underlying principle. On the other hand, Dirac
himself vacillated inconclusively between two versions of his cosmology, one
in which energy was not conserved and another in which energy was conserved\cite{r11}.\\
In any case we have here a resolution to the so called time paradox, that is
manifest irreversibility within the framework of reversible equations as alluded
to at the beginning\cite{r12}. In the words of Prigogine (loc.cit)"As is well
known, Albert Einstein often asserted, "Time is an illusion." Indeed time,
as described by the basic laws of physics, from classical Newtonian dynamics
to relativity and quantum physics, does not include any distinction between
past and future. Even today, for many physicists it is a matter of faith
that as far as the fundamental description of nature is concerned, there
is no arrow of time.\\
"Yet everywhere-in chemistry, geology, cosmology, biology, and the human
sciences-past and future play different roles. How can the arrow of time emerge
from what physics describes as a time-symmetrical world? This is the time
paradox..."\\
Infact the time of conventional physics, as argued elsewhere\cite{r6,r13}
is an approximate time, an approximation that fudges the above fluctuations.
Such a time is what may be called a stationary time, or time without time
in phraseology a la Wheeler. This is also the stationary time of Quantum
Mechanics in which the three space coordinates and time are on the same
footing as in Special Relativity, and the displacement operators represent
the energy-momenta\cite{r14}. It is the fluctuational creation of particles
which gives rise to irreversibility, and infact time itself (Cf.\cite{r15}.
It is the age old divide between being and becoming.\\
Another way of looking at this is, that in the above fluctuational cosmological
scheme, we have the analogue of the Eddington formula for time, viz.,
$$T \approx \sqrt{N} \tau ,$$
where $T$ is the age of the universe and $\tau$ the Compton time. Not only is this
relation correct, but it also provides the arrow of time.\\
We would also like to remark that the vacuum energy, as given in (\ref{e4})
is, according to latest observations, required to explain the cosmological
constant and that is, to explain the acceleration of the universe.
\section{Fluctuational Cosmology}
We now briefly describe the cosmological model referred to in Section 2\cite{r7}.
Dirac's Large Number Hypothesis (LNH) has been much written about, ever
since he spelt it out\cite{r16,r17,r18,r19,r20,r21,r22,r23,r24}. As is well known this is based on apparently
mysterious ratios of certain physical constants which coincide or show a
relationship. Let us start with,
\begin{equation}
N_1 = \frac{e^2}{Gm^2} \approx 10^{40}\label{e5}
\end{equation}
where $m$ is the pion mass, the pion being a typical elementary particle,
this being the ratio of the electromagnetic and gravitational forces, and
\begin{equation}
N_2 = \frac{cT}{l} \approx 10^{40}\label{e6}
\end{equation}
where $T$ is the age of the universe and $l$ the pion Compton wavelength.\\
In this light, the LNH can be stated as (cf.\cite{r22}).\\
"Any two of the very large dimensionless numbers occuring in nature are
connected by a simple mathematical relation, in which the coefficients are of
the order of magnitude unity"\\
An application of this to (\ref{e5}) and (\ref{e6}) means that their equality is not
accidental but rather leads immediately to, Dirac's well known relation,
\begin{equation}
G  \quad \alpha \quad T^{-1}\label{e7}
\end{equation}
Dirac's approach further leads to
\begin{equation}
R \quad \alpha \quad T^{1/3}\label{e8}
\end{equation}
which appears to be inconsistent\cite{r18,r25}.\\
Another "accidental" relation is,
\begin{equation}
m \approx (\frac{\hbar^2 H}{Gc})^{1/3}\label{e9}
\end{equation}
As observed by Weinberg (\ref{e7}), this is in a different category and is
unexplained: it relates a single cosmological parameter $H$ to constants from
microphysics.\\
In the spirit of LNH, one could also deduce that\cite{r23},
\begin{equation}
\rho \quad \alpha \quad T^{-1}\label{e10}
\end{equation}
and
\begin{equation}
\wedge \quad \alpha \quad T^{-2}\label{e11}
\end{equation}
where $\rho$ is the average density of the universe and $\wedge$ is the
cosmological constant.\\
It may be mentioned that attempts to generalise or modify the LNH have been
made (cf.eg.\cite{r25,r26}) but without gaining much further insight.\\
We now deduce equations (\ref{e7}), (\ref{e9}), (\ref{e10}) and (\ref{e11}) from
an alternative standpoint.
Moreover in place of
the troublesome equation (\ref{e9}), we will get a consistent equation. Our
starting point is the Zero Point Field (ZPF). According to QFT, this field is
secondary, while according to Stochastic Electrodynamics (SED), this field
is primary.\\
We observe that the ZPF leads to divergences in QFT\cite{r27} if no large frequency
cut off is arbitrarily prescribed, e.g. the Compton wavelength. On the
contrary, we argue that it is these fluctuations within the Compton wavelength
and in time intervals $\sim \hbar/mc^2$, which create the particles.
Thus choosing the pion again as a typical particle, we get\cite{r27,r28}
\begin{equation}
(\mbox{Energy} \quad \mbox{density} \quad \mbox{of} \quad \mbox{ZPF})
Xl^3 = mc^2\label{e12}
\end{equation}
Further as there are $N \sim 10^{80}$ such particles in the Universe, we get,
\begin{equation}
Nm = M\label{e13}
\end{equation}
where $M$ is the mass of the universe.\\
In the following we will use $N$ as the sole cosmological parameter.\\
Equating the gravitational potential energy of the pion in a three dimensional isotropic
sphere of pions of radius $R$, the radius of the universe, with the rest
energy of the pion, we can deduce the well known relation
\begin{equation}
R = \frac{GM}{c^2}\label{e14}
\end{equation}
where $M$ can be obtained from (\ref{e13}).\\
We now use the fact that the fluctuation in the particle number is of the
order $\sqrt{N}$\cite{r29,r30,r8}, while a typical time interval for the
fluctuations is $\sim \hbar/mc^2$ as seen above. (That is particles induce
more particles by fluctuations). This leads to the relation\cite{r28}
\begin{equation}
T = \frac{\hbar}{mc^2} \sqrt{N}\label{e15}
\end{equation}
where $T$ is the age of the universe, and
\begin{equation}
\frac{dR}{dt} \approx HR\label{e16}
\end{equation}
while from (\ref{e16}), we get the cosmological constant as,
\begin{equation}
\wedge \leq H^2\label{e17}
\end{equation}
where $H$ in (\ref{e16}) can be identified with the Hubble Constant, and
is given by,
\begin{equation}
H = \frac{Gm^3c}{\hbar^2}\label{e18}
\end{equation}
Equation (\ref{e14}) and (\ref{e15}) show that in this formulation, the correct
radius and age of the universe can be deduced given $N$ as the sole
cosmological or large scale parameter. Equation (\ref{e17})for $\wedge$ is
consistent and exactly agrees with an upper limit deduced for it\cite{r31}. Equation
(\ref{e18}) is identical to equation (\ref{e9}).\\
In other words, equation (\ref{e9}) is no longer a mysterious coincidence but
rather a consequence.\\
To proceed we observe that the fluctuation of $\sim \sqrt{N}$ (due to the ZPF)
leads to the empirically well known and apparently mysterious relation (\ref{e5})
\cite{r28,r29}, with $N_1 = \sqrt{N}$, whence we get,
\begin{equation}
R = \sqrt{N}l\label{e19}
\end{equation}
If we combine (\ref{e19}) and (\ref{e14}), we get,
\begin{equation}
\frac{Gm}{lc^2} = \frac{1}{\sqrt{N}}\label{e20}
\end{equation}
If we combine (\ref{e20}) and (\ref{e15}), we get Dirac's original equation
(\ref{e7}). It must be mentioned that, as argued by Dirac (cf.also ref.\cite{r32})
we treat $G$ as the variable, rather than the quantities $m, l, c \mbox{and}
\hbar$ (which we will call micro physical constants) because of their central role
in atomic (and sub atomic) physics.\\
Further, using (\ref{e20}) in (\ref{e5}), with $N_1 = \sqrt{N}$, as pointed
out before (\ref{e19}), we can see that the charge $e$ also is independant
of time or $N$. So $e$ also must be added to the list of microphysical
constants.\\
Next if we use $G$ from (\ref{e20}) in (\ref{e18}), we can see that
\begin{equation}
H = \frac{c}{l} \quad \frac{1}{\sqrt{N}} \approx
\frac{Gm^3_\pi c}{\hbar}\label{e21}
\end{equation}
Thus apart from the fact that $H$ has the same inverse time dependance on
$T$ as $G$, (\ref{e21}) shows that given the microphysical constants, and
$N$, we can deduce the Hubble Constant also, as from (\ref{e18}).\\
Use of (\ref{e17}) in (\ref{e21}) now gives equation (\ref{e11}).\\
Using (\ref{e13}) and (\ref{e19}), we can now deduce that
\begin{equation}
\rho \approx \frac{m}{l^3} \quad \frac{1}{\sqrt{N}}\label{e22}
\end{equation}
Equation {\ref{e22}) gives the equation (\ref{e10}).\\
Next (\ref{e19}) and (\ref{e15}) give,
\begin{equation}
R = cT\label{e23}
\end{equation}
The equation (\ref{e23}) differs from the troublesome Dirac dependence
(\ref{e8}).\\
Finally, we observe that using $M,G \mbox{and} H$ from the above, we get
\begin{equation}
M = \frac{c^3}{GH}\label{e24}
\end{equation}
a relation which is required in the Friedman model of the expanding universe
(and the Steady State model also (cf.refs.\cite{r24} and \cite{r25})).\\
We finally make four comments:\\
Firstly, in our model of particle production through fluctuations of the ZPF,
the equation (\ref{e15}) actually provides an arrow of time, atleast at the
cosmological scale, in terms of the particle number $N$.\\
Secondly, in the spirit of the uniform cosmic dust approximation, the newly
created particles are uniformly spread out. In practice, as the number of the
fluctuationally created particles is proportional to the square root of the
particles already present, more of the new particles are created, for example
near Galactic centres, than in empty voids, reminiscent of the jets
which are observed.\\
Thirdly, the reason why the Compton wavelength emerges as a fundamental length
has been seen in previous communications\cite{r28,r8,r33}.\\
Finally, in this model, while the mass of the universe increases as
$N \mbox{or}T^2$, the volume increases as $T^3$ so that the mean density
decreases as $T^{-1}$ (equation (\ref{e22})), unlike in the steady state cosmology.\\

\end{document}